\documentstyle[11pt,aaspp4]{article}
\def\arcs{\char'175\ ~}
\def\arcsc{\char'175 }
\def\arcm{\char'023\ ~}
\def\arcmc{\char'023\ }
\def\hub{\ifmmode H_\circ\else H$_\circ$\fi}

\journalid{337}{15 January 1989}
\articleid{11}{14}
\begin{document}

\title{Surface Brightness Profiles of Three New Dwarf Spheroidal 
Companions to M31}

\author{Nelson Caldwell}
\affil{F.L. Whipple Observatory, Smithsonian Institution, Box 97, Amado AZ 
85645}
\affil{Electronic mail: caldwell@flwo99.sao.arizona.edu}


\begin{abstract}
CCD images of three newly discovered dwarf spheroidal
companions to M31, And V, VI, and VII,
are used to extract surface brightness profiles and total magnitudes.
Using distance modulii provided by
Armandroff et al. (1998 \& 1999) and
Grebel \& Guhathakurta (1999), these galaxies are shown to have similar
luminosities to other Local Group dwarf spheroidals; And V in particular
is similar to Draco \& UMi in luminosity, i.e., among the
faintest known.  In the luminosity-metallicity
relation, And V is shown to have a significantly higher metallicity than
expected.

\end{abstract}

\keywords{ }

\section{Introduction}

In the last 20 years, 3 new dwarf spheroidal galaxies around the Milky
Way have been discovered (dwarf spheroidals, dSph, are defined here
to be the faintest of the dwarf ellipticals,  fainter than M$_{\rm V}= -14$).
These are Carina (Cannon et al. 1977), Sextans (Irwin et al. 1990)
and Sagittarius (Ibata et al. 1994).
An additional dSph, Tucana, was found unassociated with a large galaxy, but
still located within the boundaries of the Local Group (Lavery \&
Mighell 1992).
Around M31, there had been no searches and no discoveries of new
dSph's since the work of van den Bergh (1972) who found the And I, II and
III galaxies (And IV proved not be be a dSph, though its nature is
still to be settled).  This past year has seen the efforts of two groups
scanning the sky around M31, and resulting in the discovery of 3 new
M31 companions (Armandroff et al. 1998, Armandroff et al. 1999, 
Karachentsev \& Karachentseva 1999).  
All three of these dwarfs have now been the subject
of detailed study of the resolved stellar populations, resulting in 
distance and metal abundance estimates (Armandroff et al. 1998, Armandroff et al. 1999,
Hopp et al. 1999, Grebel \& Guhathakurta 1999).
HST observations are planned
for two of these as well, which will provide higher precision values for those
parameters, as well as age, and metal abundance spread estimates.  The 
basic parameter of apparent magnitude, and thus luminosity, for all three
of these galaxies are as yet unmeasured with adequate accuracy, though 
the luminosities are needed in discussions of the luminosity-metallicity
relations, for instance (Caldwell et al. 1998).  
Measuring the total magnitudes of dSph's is difficult for nearby systems
that subtend a large angle, thereby requiring the technique of counting
resolved stars down to a certain magnitude: witness the large 
errors in luminosities (typically 0.5 mag) reported for UMi
by  Caldwell et al. (1992) and Irwin \& Hatzidimitriou (1995).  
Surface photometry is possible
for systems of smaller angular size (either smaller physically or simply
more distant such as the M31 companions), and though the 
work requires a bit of care, it is straightforward and can result in
luminosity uncertainties as small as 0.1 to 0.2 mags
(Caldwell et al 1992).  Thus this paper is concerned with the task
of providing measured luminosities from surface photometry
for the new M31 companions 
And V, VI and VII.  The work
is laid out as follows.  Observations and reductions are described in the
next section, followed by the presentation of the light profiles and 
luminosities, using distances from other sources.  A few comments are made
at the end regarding how these three galaxies fit into the relations
among surface brightness, luminosity and metallicity.

And VI is also called Peg dSph and 
And VII is called Cas dSph by Grebel \& Guhathakurta (1999).  
I choose to follow the
orginal naming convention of van den Bergh, who named the companions for
the galaxy host, and not the constellations in which they were discovered
(e.g., And II is in fact in Pisces), particularly since it has already
been shown that these objects are in fact all at the distance of M31.

\section{Observational Data}
\begin{figure}[p]
\plotfiddle{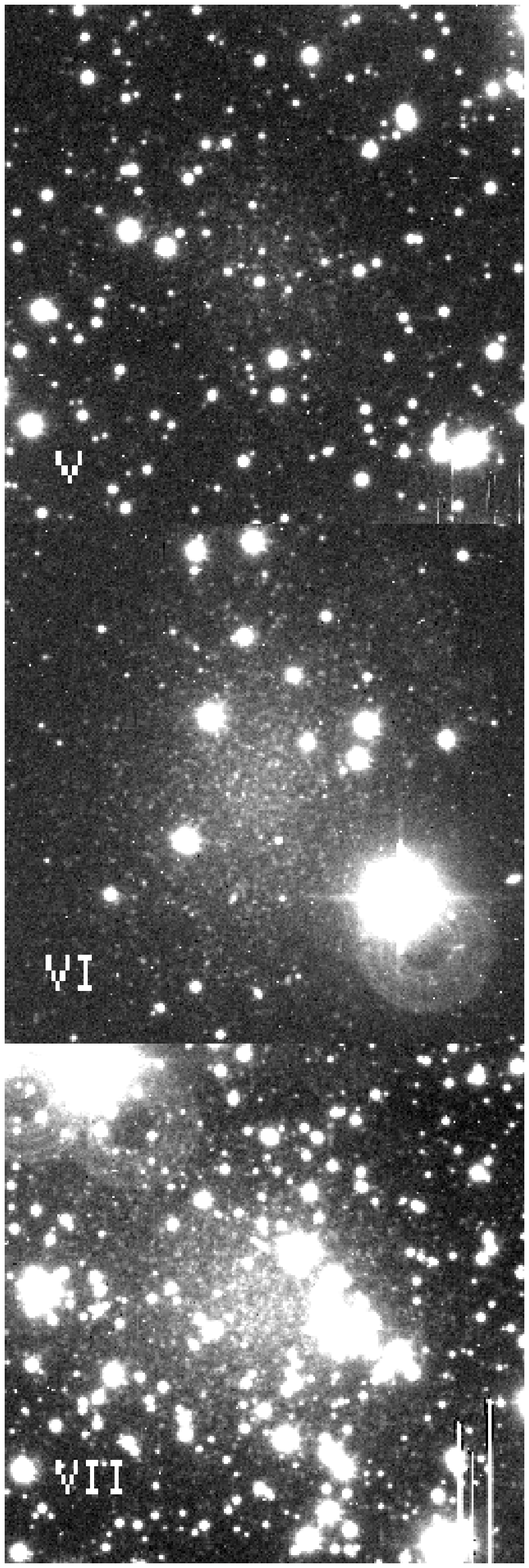}{7.3in}{0.}{60.}{60.}{-210.}{-15.}
\figcaption{1.2m images in V of the three M31 companions And V, VI and
VII. North is up; east to the left.  Field diameter for each image
is 5.7\arcm (about half of the total diameter of the images)
The intensity scale for all three images is the same.
}
\label{and_pic}
\end{figure}

The three new M31 companions were observed with the FLWO 1.2m 
telescope on Mt. Hopkins on the photometric night of
1998 Dec 11, using the ``4shooter'' CCD
camera, a mosaic of 4 2048$\times$2048 Loral CCD's.  Four 600s exposures in 
the V band were taken of And V and VI, while three such exposures were
taken of And VII.  The telescope was moved by approximately 20\arcs between
each exposure.  A dark night sky flatfield was constructed from other
data taken that night and used to flatten the dwarf frames, after debiasing.
Individual frames were shifted to a common center, and were then combined
using a simple sum, since the few cosmic rays detected would be removed
in the isophote fitting process.  The CCD's were binned on the chip 2$\times$2,
which resulted in pixel sizes of 0.67\arcsc , giving a field of 11.4\arcm per
CCD frame.  Because of the difficulty in getting different chips of a mosaic to
have the same photometric scale to better than 1\%, only the frame 
that contains the galaxy in question was used in the analysis.  After summing,
each pixel contained around 20,000 e$^-$ (except for the And VII data which
of course had 75\% of that value) due to the sky.  This value is comparable
to the sky level obtained for the And I-III data in Caldwell et al. (1992),
when account is taken of the different pixel scales, thus I should expect to
obtain the same quality of data.  The dark-sky flattened frames are flat
to $ 4\times 10^{-4} $ peak-to-valley in radial bins, which is about as good 
for the And I-III data.   Fig. \ref{and_pic} shows a montage of the 1.2m 
CCD frames on these three galaxies.

The And I-III Schmidt CCD frames of Caldwell et al. (1992)
were much larger on the sky than the 1.2m data here
(40\arcm vs 11\arcmc ), prompting some concern that the present data may not
cover sufficient area to allow the full areal extent of these new dwarfs to be
realized.  For instance, And I and II reach a surface brightness of 31 mag
arcsec$^{-2}$ at about 500\arcsc , which would be the limit in radius of 
the 1.2m 
frames.  This is certainly not a problem for the small And V galaxy, but
may affect And VI and VII in that the derived light profiles may be a bit 
steeper than they should.  The total magnitudes of these two
should not be seriously affected however, as even for And I, only 0.15 mag
of the total is found outside of 300\arcsc .  I will add 0.1 mag in 
quadrature to the uncertainties in the 
derived total magnitudes for And VI and VII to account for this concern.

The largest source of uncertainty
in the new data frames is caused by 
the presence of bright stars in the fields.  These
are masked off first automatically, by detecting all pixels above a
predetermined threshold and masking them (the level is set so that resolved
stars that are members of the dwarfs are below the level), and then by hand
masking halos of bright stars and stars within the confines of the dwarfs
that are clearly foreground.  The center of And V is somewhat contaminated
by foreground stars and thus the light profile of its central area is
more uncertain than the others.  Contrawise, the outer areas of And VII are
contaminated by very bright stars, hence its profile does not extend as far
as the others.

Uncertainties in the mean sky level of course dominate the photometric
errors, and were estimated along with the mean sky itself
by collecting data in radial bins out from the dwarf centers to the edges
of the frames.  The sky level is then the mean value acheived at large
radii, and the uncertainty is the scatter around that value.  The scatter
in the data for And VII was higher than that for the other two, because of
the bright stars mentioned, and this is reflected in the quoted uncertainty
for the derived photometric values.

An ellipse fitting program was used to collect the isophotes (Caldwell et al.
1992), once centering was set by eye. The axis ratios of And V and VII 
appeared to be circular and were fixed to those values for the isophote
collection.  The axis ratio for And VI was fit by the program, and a 
mean value for that and the position angle was determined and quoted here.
Pixels with values more than 4$\sigma $ above the mean isophotes were 
deleted, under the assumption that they were either cosmic rays or
non-member stars.  This deletion resulted in decreasing the total magnitudes 
by about 0.02 magnitudes for the dwarfs.

Photometric standards in 4 of Landolt's (1983) fields were observed during the
night, over a range of airmasses.  24 stars were used to derive the
transformation equation to V magnitudes; with a B--V color term a
standard deviation of 0.02 mag was achieved.  I thus have to 
assume a color (taken to be B--V=0.70) for the dwarf transformations
(the color term coefficient is only 0.06, so a large amount of leeway
in the actual color is allowed).

\section{Light Profiles and Luminosities}

\begin{figure}[p]
\plotfiddle{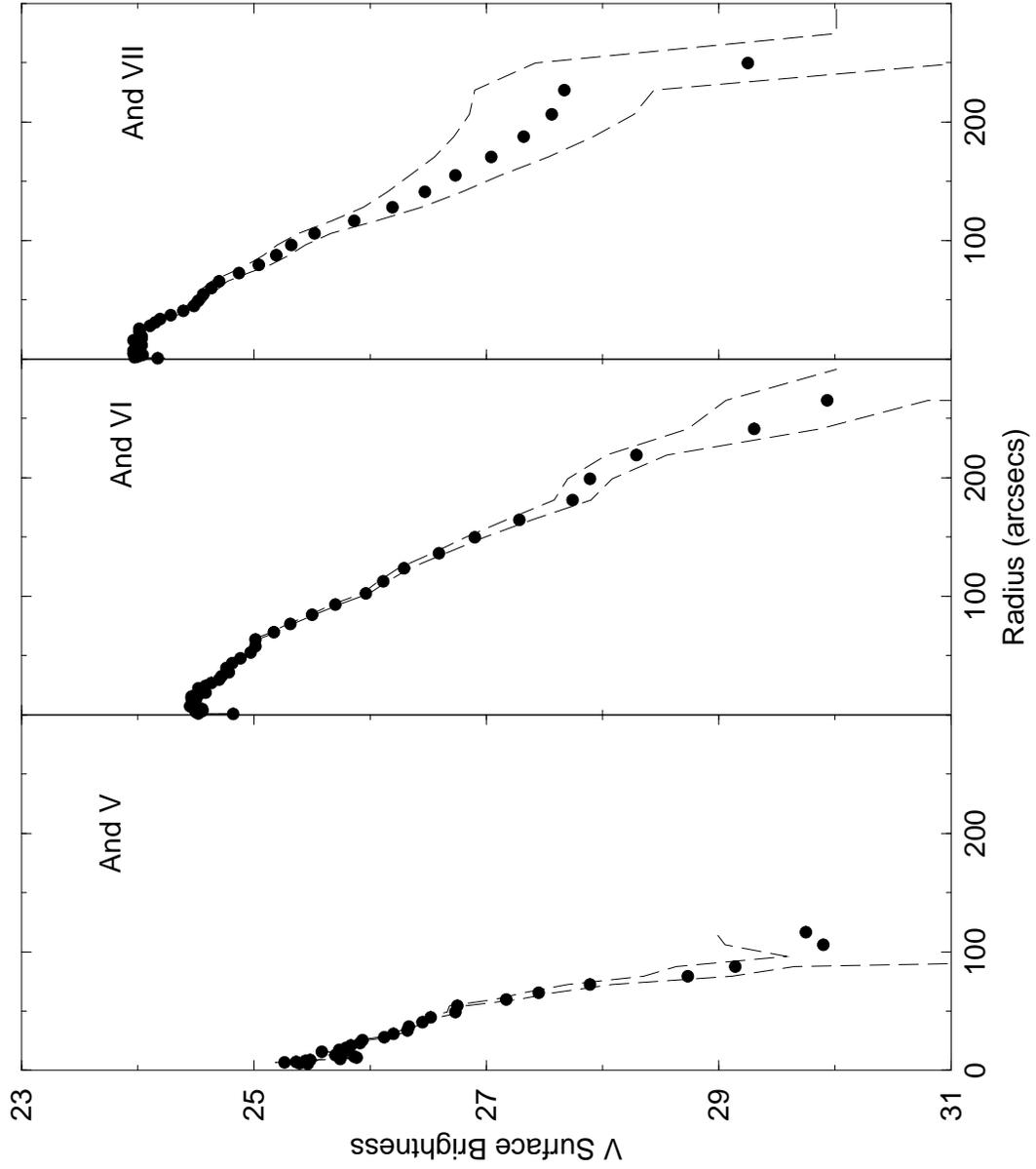}{7.1in}{0.}{78.}{78.}{-233.}{0.}
\caption{Radial light profiles of the three M31 dwarfs.  Surface
brightness in V mags arcsec$^{-2}$ is plotted against the 
geometric mean of the semi-axes in arcsecs.  Dashed lines
show 1$\sigma $ error bars in the surface brightnesses.}
\label{all.pro}
\end{figure}

Fig. \ref{all.pro} shows the derived
radial light profiles for these three galaxies.  Errors are indicated
by the dotted lines above and below the surface brightness points, and
are derived from photon statistics, readout noise
and the uncertainty in the mean sky level.  The profile for And VI
matches well that done for the galaxy by Hopp et al. (1998). The differences
are of the order of 0.05 mag in surface brightness, and there appears to 
be no systematic offset.

And V as expected is much 
smaller than the other two, and of lower surface central brightness.
The central surface brightness for And VII is relatively high, is
higher than any of And I-III, and indeed is next to 
Leo I and Fornax in this regard.  Its brightness
makes its delayed discovery more difficult to understand, aside from the
obvious fact that no one looked in detail on sky 
survey plates in its area until the
discovery group did (Karachentsev \& Karachentseva 1999).

A number of important quantities can be derived from these light profiles;
these are listed in Table \ref{table1}.  Radii are geometric means
of the semi-axes, which only matters for And VI since the other two are
taken to be circular in projection.  The total apparent magnitude
is calculated by summing up the profiles, to the point in radius where the
isophote level goes to zero.  The quoted uncertainties include the 0.1 mag
value discussed in section 2 for And VI and VII.  Absolute magnitudes
are calculated using the distance modulii from Armandroff et al. (1998, 1999)
and Grebel \& Guhathakurta (1999).   Errors are the 
quadratic mean of those of the apparent magnitudes and the distance
modulii.  Reddenings listed are from Burstein \& Heiles (1982) 
or Schlegel et al. (1998) for
And V and VI (used by Armandroff et al. 1998 and 1999 
in determining metallicities via the color of 
the giant branches), and from Grebel \& Guhathakurta (1999)
for And VII, a combination of reddenings from Burstein \& Heiles and
Schlegel et al.  The Schlegel et al. new Galactic reddening 
description gives a somewhat smaller reddening than
Burstein \& Heiles for And V, and a somewhat
larger reddening for And VI.

And V has an unusually faint
total magnitude, of M$_{\rm V}$ = --9.1, which puts it in the realm
of the faintest Galactic dSph's, Draco, Uminor, and Carina (and in fact
And V now has the most accurate measure of M$_{\rm V}$ for any of these).
Hopp et al. (1999) had quoted an M$_{\rm V}$ = --10.4 for And VI, but this
value is in fact only the magnitude within a V surface brightness of 25.5 
mag arcsec$^{\rm -2}$ as stated in their text, and not the total magnitude
which as shown here is nearly 1 magnitude brighter, at M$_{\rm V}$ = --11.3.
The M$_{\rm V}$  = --12.0 for And VII places it next in luminosity
just below Fornax and Sagittarius in the Local Group.  

Central surface brightnesses are estimated directly from the light
profiles; the values in Table \ref{table1} are corrected for
the quoted extinctions.  As mentioned, the foreground stars near the 
center of 
And V make its central surface brightness more uncertain than is the case
for the other two.  The value here of V$_{\rm 0}$ = 24.8 is higher than
found in Armandroff et al. (1998) by 0.4 mag.  This appears to be due to
the fact that those authors averaged the data over the central 20\arcsc .  
Averaging the present And V data over that region results in a similar value
to that of Armandroff et al.
The value for
And VI of V$_{\rm 0} = 24.36 \pm 0.05$ matches well that of $24.2$ shown
in Fig 2 of Hopp et al. (1999). (This is not the value quoted in their text,
which refers to their exponential fit.)

Sersic profiles (I=I$_{\rm 0}e^{-(r/r_{\rm 0})^n}$,
Sersic 1968), which have an additional free parameter $n$ over an
exponential ($n$=1), were fit to the data. 
Table \ref{table1} shows the derived values (S$_{\rm 0} =
-2.5 \rm{log}_{10}~\rm{I}_{\rm 0} $ \ and is corrected for extinction).  
As is typical for the faintest dSph's, And V has a profile much steeper than 
an exponential.

The empirical core radii refer to the radius at which the local surface
brightness has fallen to half the central value.  These were measured
directly off of the profiles.  Effective radii refer to the radius which
contains half the total light, and were likewise obtained directly from 
the profiles.  And VI and VII are typical sized dSph's for the Local 
Group.  And V is among the smallest, its R$_{\rm e}$ = 145 pc being 
smaller than UMi (190, Caldwell et al. 1992) and 
Carina (190), and close to that of Draco
(133, Irwin \& Hatzidimitriou 1995 for the last two). 
Again, the value for And V is more accurate than those for the
Galactic dwarfs.

\section{Comments}
\begin{figure}[p]
\plotfiddle{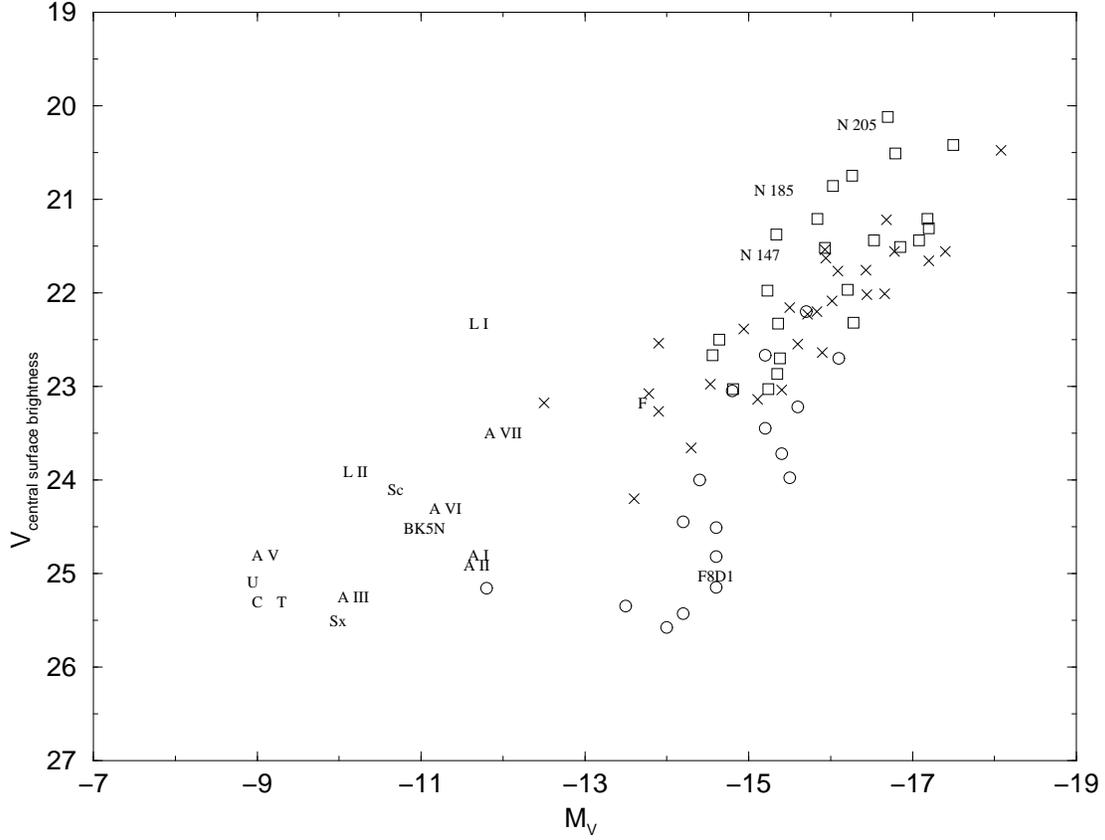}{3.0in}{-90.}{70.}{70.}{-270.}{400.}
\caption{Relation between central surface brightness and M$_{\rm V}$
for dwarf ellipticals.  Crosses and open squares represent Virgo and
Fornax dE's; circles represent the large, low surface brightness galaxies
found in Virgo by Impey et al.\ (1988); Local Group dE's are shown
with abbreviations for their names; and the M81 group dwarfs, BK5N
and F8D1, are shown with their names.  Draco is not plotted because 
neither its central surface brightness nor its luminosity are as
well known as the other galaxies plotted here. }
\label{Mv_vo}
\end{figure}

How do these three new M31 dwarfs fit it with the family of dwarf
ellipticals?  Fig. \ref{Mv_vo} shows the relation between luminosity
and central surface brightness, a diagram discussed in Caldwell 
et al. (1998).  The new M31 dwarfs occupy the same area as the
other Local Group dwarfs occupy.  And V as noted above is among the
faintest known dSph's.  And VII appears to have high surface brightness
for its luminosity, though not nearly as strong a case as Leo I.
As a group, the 6 known M31 companions are very similar to the 8 known
Galactic dSph's.

\begin{figure}[p]
\plotfiddle{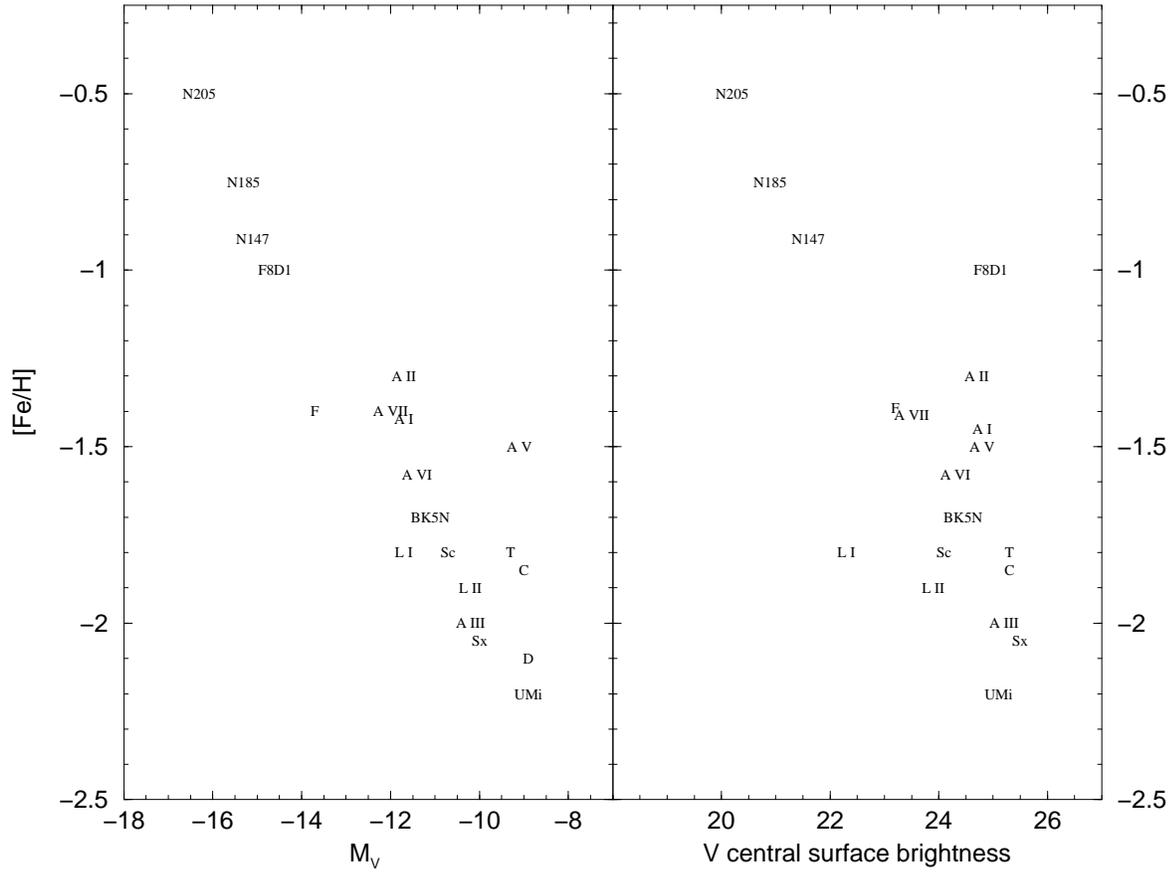}{4.0in}{-90.}{65.}{65.}{-250.}{400.}
\caption{Relation between metallicity ([Fe/H]) and M$_{\rm V}$;
and metallicity and central surface brightness.  Additional 
new metallicities shown in this plot (changes over what 
was shown in Caldwell et al. 1998) are those 
of NGC~147 (Han et al. 1997) and NGC~185 and 205 (Geisler et al. 1999).}
\label{feh_l}
\end{figure}

The detailed work on the resolved stellar populations 
of And V and VI by Armandroff et al. (1998 and 1999), and on And VI and VII by
Grebel \& Guhathakurta (1999) have provided good estimates of the mean
metallicities of these dwarfs from the colors of the giant branches, 
and thus allow a further comparison of
these dwarfs with others to be made.  The relation between luminosity and
metallicity for old stellar populations is now well studied, if not yet
completely understood (Mould, Kristian, and Da Costa 1983, Caldwell et al. 
1998, Mateo 1998).  There appears to be a similar relation
for star forming dwarfs (Aaronson 1986, Skillman et al. 1989,
Richer \& McCall 1995); comparing the two
requires knowing [O/Fe] in most cases because the kinds of metallicities 
measured for the two types of dwarfs
are different ([Fe/H] of stars for dSph's; [O/H] for star forming dwarfs).
Fig. \ref{feh_l} shows the new M31 dwarfs in relation to other
dwarfs whose metallicities have been measured in the same way, in the
luminosity-metallicity plane, and in the surface brightness-metallicity plane.  
For And V and VI,
the [Fe/H] values (--1.5 \& --1.58, respectively) 
of Armandroff et al. (1998 \& 1999) have been chosen, while for And VII the [Fe/H]= --1.4
comes from Grebel \& Guhathakurta (1999).  

And VI and VII appear to have [Fe/H] values expected for their luminosities,
but And V has a metallicity high for its very low luminosity.  If the lower
Schlegel et al. (1998) reddening is adopted for And V, it's metallicity of
course increases, making the discrepancy even larger (the other faint dSph's
are not much affected by changes in adopted reddening).
The range
in [Fe/H] for galaxies with M$_{\rm V} \sim -9 $ is about 0.7.  This is not
as large as the range in [Fe/H] for galaxies with low surface brightness
(1.2), so it still seems clear that luminosity is the dictating factor in
a dwarf's mean metallicity.
Clearly, a larger sample of the new, very faint dSph, either in the Local Group
or in nearby groups, which also would be studied in detail, 
would be helpful in this matter.

Mateo (1998) interprets the luminosity-metallicity relation as being bi-modal,
based on the fact that the Local Group dSph galaxies brighter
than M$_{\rm V} \sim -14 $ had lower metallicities than an extrapolation
of the relation determined by all fainter galaxies would predict.
New metallicities for those brighter galaxies from Han et al. (1997) 
and Geisler et al. (1999) are shown in Fig \ref{feh_l}, indicating
that the relation defined by the lower galaxies extends well into the
brighter galaxy region, thus negating the need for a bimodal interpretation.
The position of And V in this
relation, with its metallicity offset of $\sim +0.5$ dex from the other
galaxies in its luminosity bin, however, shows that a large scatter 
exists for the lower luminosity galaxies. It may be that the scatter
at all luminosities in the relation is  larger than currently evident -
more data on luminous galaxies would be helpful.

Finally, And V's high metallicity for its luminosity  may imply an even  
deeper potential well than is the case for say, Draco and Umi
(Pryor \& Kormendy 1990).  If so, its
stellar velocity dispersion would thus be higher than those two, and so also 
its  M/L ratio.  Securing stellar kinematics of this galaxy then takes on 
an even larger importance, for the the similarity of dark 
matter halo densities for low mass halos could be investigated
(Navarro et al. 1997).

\acknowledgements

I thank T. Armandroff and G. Jacoby for useful discussions and 
exchange of data prior to publication.  Ed Olszewski kindly provided a
critical reading of the text.

\clearpage

%
%



%


\clearpage

%
%

\begin{deluxetable}{lrrr}
\tablenum{1}
\tablecolumns{4}
\tablewidth{0pc}
\tablecaption{Basic Data for And V, VI, and VII \label{table1}
}
\tablehead{\colhead{Parameter}& \colhead{And V} &\colhead{And VI}
&\colhead{And VII}
}
\startdata
RA$_{\rm J2000}$ &01:10:17.1 & 23:51:46.3 & 23:26:31\nl 
Dec$_{\rm J2000}$ & 47:37:41 & 24:34:57 & 50:41:31\nl
(m--M)$_0$ & $ 24.55 \pm 0.12$ &  $24.45 \pm 0.1$ & $ 24.4 \pm 0.2 $ \nl
V$_{\rm tot}$&$15.92 \pm 0.14$& $13.30 \pm 0.12$& $12.90\pm 0.27$ \nl
M$_{\rm V}$& $-9.1\pm 0.2$ & $-11.3 \pm 0.2 $ & $-12.0 \pm 0.3 $\nl
V$_{\rm 0}$ & $24.8 \pm 0.20 $& $24.31 \pm 0.05$&$ 23.47 \pm 0.05$\nl
r$_{\rm c}$(pc)&  110& 286& 240 \nl
R$_{\rm e}$(arcsec) &37& 84&80 \nl
R$_{\rm e}$(pc) & 145& 316& 295 \nl
S$_{\rm 0}$ &$25.01 \pm 0.06$&$24.20 \pm 0.03$&$23.34 \pm 0.03$ \nl
r$_{\rm 0}$(arcsec)& $45 \pm 2 $& $ 82 \pm 2$& $75 \pm 2$ \nl
n  & $1.7 \pm 0.16$& $1.38 \pm 0.04$& $1.32 \pm 0.04$ \nl
1-b/a&0& 0.23& 0 \nl
PA  & \nodata& 160& \nodata \nl
A$_{\rm V} $& 0.50 & 0.19 & 0.53 \nl
\enddata
\tablenotetext{}{Coordinates and distance modulus for And V and And VI
from Armandroff et al. (1998, 1999); those for
And VII from Grebel \& Guhathakurta (1999). Surface brightness values
are dereddened.}
\end{deluxetable}
\end{document}